\documentclass[showpacs,prl,aps,twocolumn]{revtex4}
\usepackage{graphicx}
\usepackage{amsmath}
\usepackage{bm}
\def\gapx{\lower 2pt \hbox{$\buildrel>\over{\scriptstyle{\sim}}$}}
\def\lapx{\lower 2pt \hbox{$\buildrel<\over{\scriptstyle{\sim}}$}}
\begin{document}
\widetext
\title
{Phase diagram of  anisotropic boson $t$-$J$ model }
\author{Massimo Boninsegni$^{1}$ and Nikolay V. Prokof'ev$^{2,3,4}$}
\affiliation {
$^{1}$Department of Physics, University of Alberta, Edmonton, Alberta, Canada T6G 2J1\\
$^{2}$Department of Physics, University of Massachusetts,
Amherst, MA 01003, USA\\
$^{3}$Theoretische Physik, ETH Z\"{u}rich, CH-8093 Z\"{u}rich, Switzerland \\
$^{4}$Russian Research Center
``Kurchatov Institute'', 123182 Moscow, Russia
}
\date{\today}
\begin{abstract}
{We have studied by Quantum Monte Carlo simulations the low temperature
phase diagram of a mixture of isotopic, hard core bosons, described by
the $t$-$J_z$-$J_\perp$ model, with $J_\perp$=$\alpha J_z$. Separation
of superfluid hole-rich and insulating, antiferromagnetically ordered
hole-free phases is observed at sufficiently low hole density, for any
$\alpha < 1$. A two-component checkerboard supersolid phase is not observed.
The experimental relevance and possible broader implications of these
findings are discussed.
 }
\end{abstract}

\maketitle
Impressive scientific and technological advances in trapping cold
atoms in optical lattices (OL) \cite{jaksch98,jaksch05,greiner02,review} render it
now feasible to create in the laboratory remarkably close
experimental realizations of model many-body systems long thought of
as of mostly academic interest. An example is the isotopic
two-component Bose mixture, providing a rich playground for
many-body physics due to the various phases that it is expected to
display, including a number of physically distinct superfluid phases
\cite{proko04a,kuklov}. The ongoing experimental investigation of this
system justifies the theoretical exploration of its phase diagram,
not only to help in the interpretation of experimental data, but
also for the more general purpose of guiding the search for novel
phases of strongly correlated quantum many-body systems.

In this work, we model a mixture of two different species of hard core
bosons (the limit of large on-site repulsion is assumed 
for both like and unlike bosons) of equal masses,  via the two-dimensional (2D) boson
$t$-$J_z$-$J_\perp$ model \cite{boninsegni01}:
\begin{eqnarray} \label{two}\nonumber
&{\hat H}& = - t \ \sum_{\langle ij\rangle} \biggl( {\hat a}^\dagger_i
{\hat a}_j + {\hat b}^\dagger_i {\hat b}_j + h. c.  \biggr)  \\
&-& {1\over 2}  \biggl [ J_z \biggl
({\hat n}_i {\hat m}_j + {\hat m}_i {\hat n}_j \biggr ) +  J_\perp \biggl(
{\hat a}^\dagger_i {\hat a}_j {\hat b}^\dagger_j {\hat b}_i + h. c.
\biggr )  \biggl ]\;\;\;
\end{eqnarray}
A square lattice of $N$=$L$$\times$$L$ sites is assumed, with periodic boundary conditions (PBC). Two species ($A$ and $B$) of bosons of equal masses are defined,
for which
${\hat a}^\dagger_i$, ${\hat b}^\dagger_i$, are creation operators,
whereas ${\hat n}_i={\hat a}^\dagger_i
{\hat a}_i$, ${\hat m}_i={\hat b}^\dagger_i {\hat b}_i$ are
number operators.
The sum in (\ref {two}) runs over all pairs of nearest-neighboring (NN) sites.

The Hamiltonian (\ref{two}) is defined in the subspace
in which no double occupation of sites is possible. The parameters of the model, namely $t$, $J_z$ and $J_\perp$ are all non-negative; henceforth, we shall take $t$ to be our energy scale, and set it equal to one. We set $J_z = J$ and $J_\perp = \alpha J$, $\alpha$ expressing the anisotropy between the ``antiferromagnetic" coupling $J$, represented by the second term in (\ref{two}), and the ``ferromagnetic" exchange
coupling $J_\perp$, represented by the last term in (\ref{two}).
In this work, we focus on the parameter region in which both $\alpha$
and $J$ are less than 1.
The hole density is defined as
$h \equiv 1 - ({N_{A}+N_{B}})/{N}$, where $N_{A}$ ($N_{B}$) is the number of particles of species $A$ ($B$). All throughout, we assume $N_{A}=N_{B}\le N/2$, i.e., the system has no net ``magnetization''.  The isotropic version of (\ref{two}) (i.e., with $\alpha=1$), can be derived from the two-component isotopic  Bose Hubbard model in the limit of large on-site repulsion and small hole concentration \cite{boninsegni01,boninsegni02,duan03}; in an optical lattice, anisotropy could arise from additional longer-ranged (e.g., dipolar) interactions among particles. At exactly half filling (i.e., no holes),
(\ref{two}) can also be cast in the spin language; for example, it is isomorphic to a spin-1 antiferromagnetic  Heisenberg model with uniaxial single-ion anisotropy, possibly relevant to  some magnetic systems \cite{cristian}.

The fermion counterpart of (\ref{two}) with $\alpha$=1 is known as the $t$-$J$ model \cite{jorge,anderson87,zhang88}, and has been the subject of a wealth of theoretical work, because of its posited connection to high-temperature superconductivity (HTS).  In spite of an enormous effort now spanning almost two decades the phase diagram of the $t$-$J$ model remains relatively poorly understood. Basic questions, such as the presence of a superconducting ground state, are yet largely unanswered, essentially due to the lack of  a sufficiently robust theoretical method for strongly correlated fermions.
On the other hand, the case of  Bose statistics (\ref{two}) can be studied by Quantum Monte Carlo (QMC) simulations, yielding essentially exact numerical results.
For this reason, some studies have used (\ref{two}) as a starting point to investigate physical effects such as stripe formation, also believed to be relevant to HTS \cite{smakov04}.

The realization of strongly correlated models such as (\ref{two}) in OL is the goal of a current, intense
experimental and theoretical effort. It seems therefore worthwhile to provide quantitative, reliable theoretical information on phase diagrams of these models, especially to guide the experimental search for novel phases of matter.

In this work, we study the low temperature phase diagram of $\hat H$ as a function of the hole density $h$, for $\alpha \le 1$ and different values of $J$. Specifically, we consider the physically realistic region $J$ $<$ 1 for which the isotropic version of (\ref{two}) displays no phase separation, as shown in Refs. \onlinecite{boninsegni01,boninsegni02}.  Of particular interest is the effect of  anisotropy  on the phase diagram of (\ref{two}), chiefly with regard to phase separation, superfluidity, and the possible presence of a two-component checkerboard supersolid phase (SPS).  We use the Worm Algorithm (WA) in the lattice path-integral representation \cite{worm} to compute thermal expectation values of physical operators; the calculation requires a relatively straightforward extension of the WA, to allow for the simulation of physical processes associated with the exchange term of (\ref{two}).

In the absence of holes (i.e., $h=0$),  the Hamiltonian (\ref{two}) does not differentiate between Bose and Fermi statistics; for, upon performing a straightforward basis transformation (possible on a
bipartite lattice), the SU(2) symmetry of the Bose Hamiltonian can be restored.
For $\alpha < 1$, the system features a transition (at finite temperature $T_N$)  to an antiferromagnetically ordered state, in which a particle of type $A (B)$ is preferentially surrounded by particles of type $B (A)$ \cite{tognetti}. At $\alpha=1$, it is $T_N=0$, but the ground state is still ordered \cite{manousakis}. For $\alpha>1$, the system is superfluid (SF)
below the Berezinskii-Kosterlitz-Thouless (BKT) transition temperature $T_{BKT}$, but with no net flow of matter, as the flow of one component is exactly compensated by counterflow of the other \cite{SCF}.

The presence of mobile holes is generally expected to result in a reduction of the antiferromagnetic order, as holes scramble it with their motion, as well as to give rise to a SF phase of holes at low temperature; of interest is the possible existence of a {\it supersolid} phase, in which both types of orders may coexist in a single homogeneous phase.

The occurrence of checkerboard  or antiferromagnetic order can be ascertained by numerical simulations of the staggered density order parameter $\chi(h)=\sqrt{S(\pi,\pi)}$ where  $S({\bf q})=\langle \hat \rho_{\bf q} \hat \rho_{-{\bf q}}\rangle$, with
\begin{equation}
\hat\rho_{\bf q}={1\over N}\sum_i\ e^{i{\bf q}\cdot{\bf r}_i}\  (\hat n_{i}-\hat m_{i}) \;,
\end{equation}
and where ${\bf r}_i$ is the position of the $i$th lattice site, and $\langle ... \rangle$ stands for thermal expectation value. The SF density $\rho_S (h)$ of the fluid of holes is computed using the usual ``winding number" estimator.
\begin{figure}[h]
\centerline{\includegraphics[height=3.2in, angle=-90]{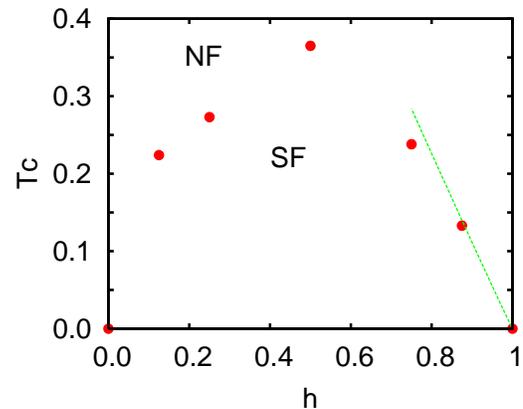}}
\caption{Superfluid transition temperature of the uniform hole gas in the model (\ref {two}) with $\alpha=1$ (i.e., isotropic), as a function of hole density $h$. Filled circles show numerical estimates of $T_{\rm c}$ (in units of $t$), determined as explained in the text. Dotted line at high $h$ is the theoretical behavior predicted by the theory of the dilute Bose gas \cite{WIBG2D}. Here, $J=0.4$.}
\label{fig2}
\end{figure}

We discuss the isotropic ($\alpha$=1) case first. As shown in Refs. \cite{boninsegni01,boninsegni02}, the system features a homogeneous ground state, for any hole concentration, for $J$  $\lapx$ 1.5. The AF order that exists in the undoped system at $T$=0, is suppressed by an arbitrarily small hole concentration.
The underlying mechanism is simple: holes break the SU(2) symmetry of the undoped 
Hamiltonian in favor of the XY-plane and orient the order parameter. 
 We have explicitly verified this conclusion by performing finite-size scaling analysis of $\chi(h)$. This physical result is analogous to that observed for a system of lattice hard core bosons with a nearest-neighbor repulsive interaction potential of strength $V=2t$, for which doping away from half-filling destroys checkerboard order \cite{batrouni}. This points to a significant difference between model (\ref{two}) with Fermi and Bose statistics. In the case of the fermion $t$-$J$ model the AF is reduced but not completely eliminated by hole doping \cite{spanu} since the SU(2) spin symmetry is preserved.

The ensuing, uniform hole gas is SF at $T$=0, and features a BKT transition to a normal fluid (NF) at finite temperature. The transition temperature $T_{\rm c}$ can be obtained using
the well known renormalization flow and the universal jump of the
superfluid density, $\rho_s$, at $T_c$ \cite{triang}. Results are shown in Fig. \ref{fig2} for the case $J$=0.4.
\begin{figure}[h]
\centerline{\includegraphics[height=3.2in, angle=-90]{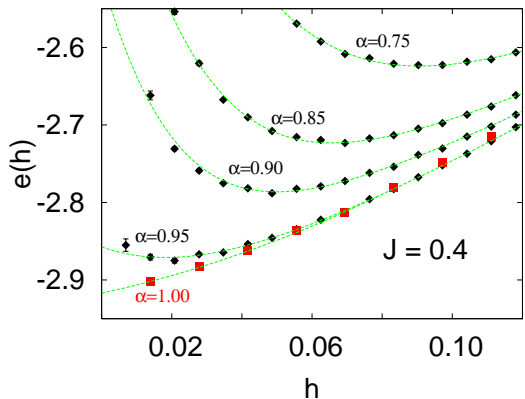}}
\caption{Ground state energy per hole $e(h)$, defined in (\ref{epsh}), for the Hamiltonian (\ref {two}) as a function of the hole density $h$, for different values of the anisotropy parameter $\alpha$ (diamonds). Boxes show estimates of $e(h)$ for the isotropic ($\alpha$=1) case. Dotted lines are polynomial fits to the data.  Numerical calculations were carried out on a 12$\times$12 lattice. Here, the temperature $T$=0.025 $t$ and $J=0.4$.}
\label{fig1}
\end{figure}

A richer phase diagram occurs in the anisotropic ($\alpha < 1$ case).
The ground state energy per hole is defined as \cite{emery90,notegr}
\begin{equation}\label{epsh}
e(h) = {E(h)-E(0)\over Nh}
\end{equation}
where $E(h)$ is the total energy of the system in the presence of $Nh$ holes. A minimum of $e(h)$ at a finite hole density $h_{cr}$,  signals the separation of the system into two phases at $h < h_{cr}$, one with no holes, and the other with hole density $h_{cr}$.

Fig. \ref{fig1} shows $e(h)$ as a function of hole density and for different values of the anisotropy parameter $\alpha$. The value of $J$ is 0.4. Our numerical results show a minimum for $e(h)$ at a finite hole concentration $h_{cr}(\alpha)$, for any $\alpha < 1$; that is, the system separates into hole-rich and hole-free phases for hole doping below $h_{cr}(\alpha)$ (with $h_{cr}(\alpha)\to 0 $ as $\alpha \to 1$).
\begin{figure}[h]
\centerline{\includegraphics[height=3.45in, angle=-90]{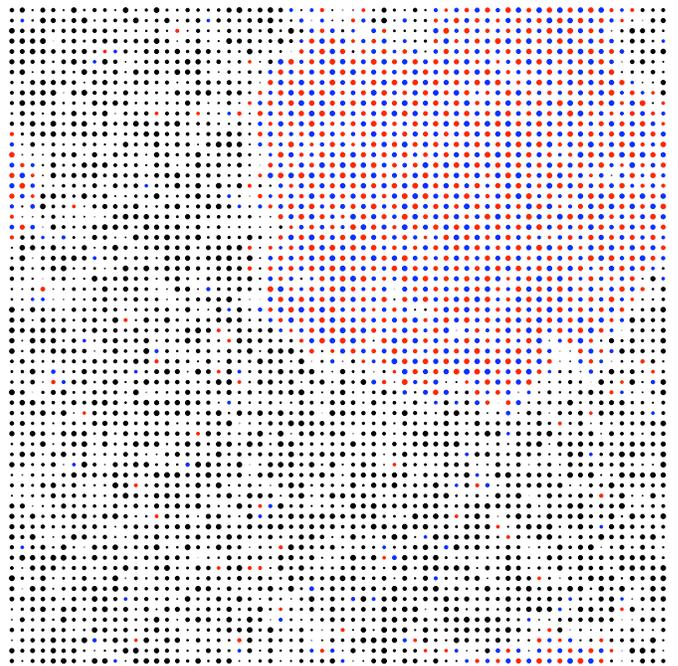}}
\caption{Snapshot of an instantaneous ($\beta$-averaged) configuration of the system of a $64^2$-site lattice. Color coding is as follows: sites that are drawn in red (blue) are occupied by a particle of type A, with a greater probability the bigger the size of the circle; sites that are drawn in black are occupied by holes and both types of particles with similar probability, the smaller the circle, the more likely for the site to be empty. Here, $J$=0.4, $J_\perp$=0.3, and $h$=0.0586 at $T$=0.025 $t$.
}
\label{fig3}
\end{figure}
The hole-free phase features AF order (and obviously no hole-based superfluidity), whereas the hole-rich phase is SF, but not antiferromagnetically ordered. The occurrence of PS in the anisotropic model can be understood based on the ``string" picture \cite{brinkman}. In the anisotropic model, a hole leaves
behind, in its motion, a string of bosons of either species, displaced by
one lattice site. Thus,  separation of the system
in hole-rich and hole-free phases becomes energetically advantageous at
low hole density, as a way to limit the damage caused by the holes to the
antiferromagnetic order. In the $\alpha\to 1$ limit, however, quantum fluctuations
associated with the $J_\perp$ term of (\ref {two}) ``mend'' the damage
due the hole motion, restoring local order \cite{boninsegni}.
These considerations are clearly independent of quantum statistics, i.e., the ought to
apply to the fermion models as well.

The separation of the two phases can be visually observed by examining configurations generated at low $T$ by our algorithm on lattices of sufficiently large size, e.g., $N=4096$ sites (see Fig. \ref{fig3}). We have consistently observed this effect for $J < 1$, and for as little as 1\% anisotropy. Based on this numerical evidence, we argue that PS will occur in the $T\to 0$ limit, at sufficiently low hole concentration, for {\it arbitrarily weak} anisotropy.

At finite temperature, entropy favors  mixing of the two phases and the occurrence of a homogeneous phase. Because both SF and AF transition temperatures are finite, an obvious question is whether a homogeneous phase featuring both types of order, namely a ``two-component checkerboard supersolid (i.e, a superfluid gas of holes inside a checkerboard quantum antiferromagnet formed by the two components), may exist at finite temperature.
\begin{figure}[h]
\centerline{\includegraphics[height=2.8in, angle=0]{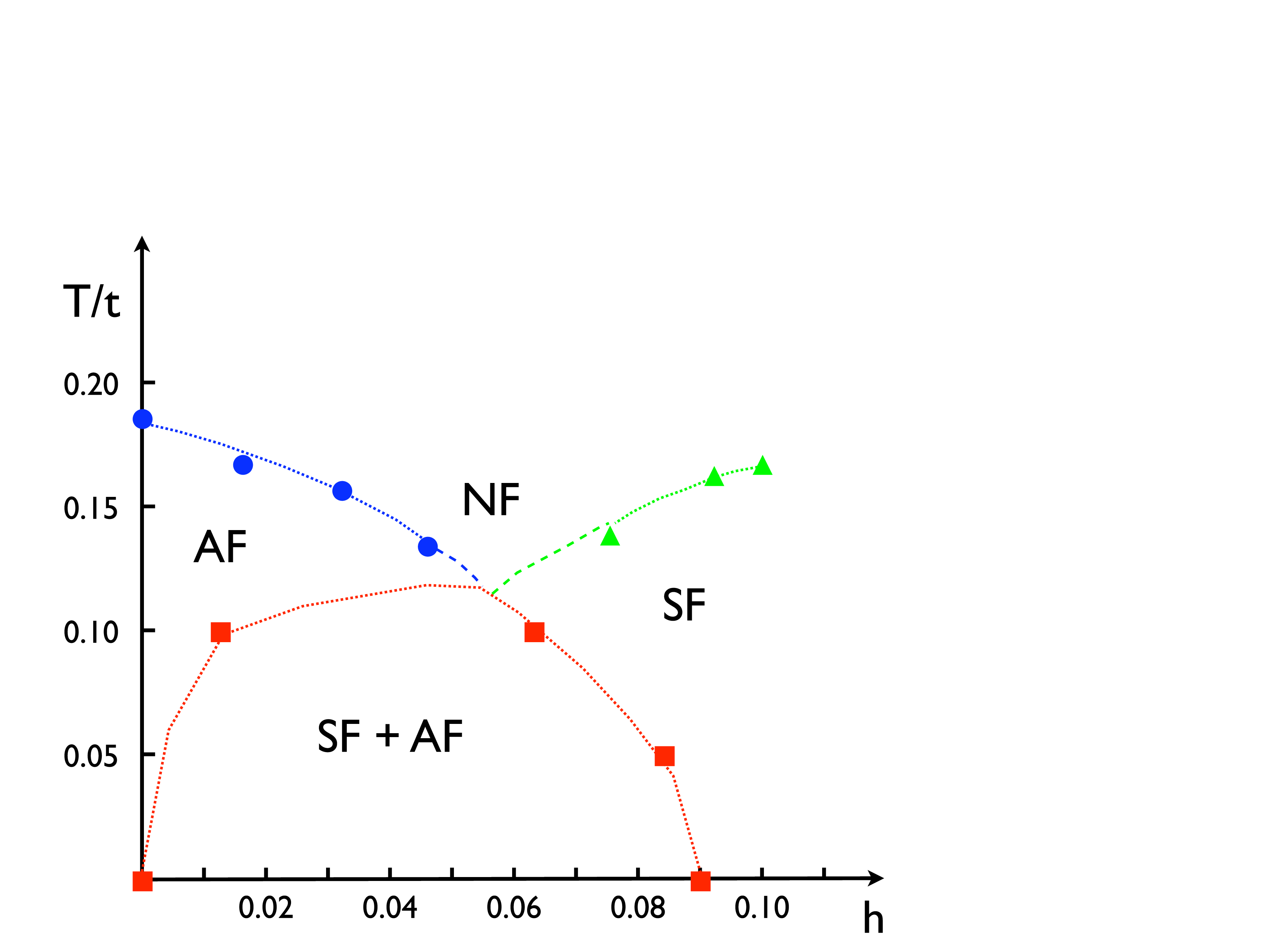}}
\caption{
Computed phase diagram of the boson $t$-$J_z$-$J_\perp$ model on the square lattice. Horizontal axis shows hole concentration, vertical temperature. Figure shows actual Monte Carlo data obtained for $J_\perp =0.3, \  J=0.4$, but the same schematic phase diagram is obtained for all values of the model parameters considered here. Circles represent the normal-to-antiferromagnetic transition, triangles the normal to superfluid, whereas boxes show the boundary of the region in which coexistence of superfluid and antiferromagnetic phases is observed. Dashed lines are only meant as a guide to the eye. Statistical errors are comparable to the sizes of the symbols.
}
\label{fig4}
\end{figure}

Our simulations have not yielded any evidence of such a checkerboard supersolid phase. On raising the temperature, the system evolves into either a non-superfluid antiferromagnet, or into a SF with no antiferromagnetic order. We find that the
region where all transition lines come close to each other is extremely hard
to study since all standard finite scaling techniques fail. In Fig.~\ref{fig4}
we sketch the simplest phase diagram consistent with our data, also the one which we find
most plausible; in this scenario, NF-SF and NF-AF lines meet the coexistence dome at one point, close to its maximum. However, at this time we can not exclude other scenarios. For example,
the AF line may feature a tri-critical point, and the BKT transition
may be terminated at the first-order AFM line. The possible crossing of the
second order AF and BKT lines, giving rise to a small region of existence of a two-component checkerboard supersolid sitting over the phase separation dome has been throughly investigated, but not observed in any of our simulations.

There are obviously interesting similarities between the phase diagram schematically represented in Fig. \ref{fig4} and the basic phase diagram of HTS. In the case of the model investigated here, anisotropy is crucial in stabilizing the AF phase, which disappears upon doping in the fully isotropic model, as shown above. Anisotropy might also play a role in shaping the phase diagram of the fermion counterpart of (\ref {two}), deemed relevant to HTS, as assuming a stronger AF coupling in the $z$ direction may be physically justified by considerations of interplane exchange coupling \cite{birg}. As mentioned above, however, in the case of Fermi statistics AF order is not completely destroyed by the injection of mobile holes.

The interesting interplay of phases shown in the phase diagram of Fig. \ref{fig4} suggests that an experimental system described by (\ref{two}) may be worth investigating in OL.
This work was supported in part by the
National Science Foundation under Grant No. PHY-0653183 and by the Natural Science
and Engineering Research Council of Canada under research grant G121210893. The authors gratefully acknowledge hospitality and support of the INFM-BEC Research and Development Center, Universita' degli Studi di Trento, Italy.


\begin{thebibliography}{99}
\bibitem{notegr}
By ``ground state" we mean that estimates are obtained at a temperature $T$ sufficiently low that they no longer change, within our statistical precision, upon lowering $T$ any further. In the case of the results shown in Fig. \ref{fig1}, it is $T$=0.025 $t$.
\end{thebibliography}
\end{document}